\documentclass[prl,aps,amssymb,twocolumn,showpacs,superscriptaddress]{revtex4}
\usepackage{graphicx}
\usepackage{amsmath}

\begin{document}

\title{Electron-photon correlations and the third moment of quantum noise}

\author{Julien Gabelli}
\affiliation{Laboratoire de Physique des Solides, Univ. Paris-Sud, CNRS, UMR 8502, F-91405 Orsay Cedex, France}
\author{Lafe Spietz}
\affiliation{National Institute of Standards and Technology, Boulder, Colorado 80305, USA}
\author{Jose Aumentado}
\affiliation{National Institute of Standards and Technology, Boulder, Colorado 80305, USA}
\author{Bertrand Reulet}
\affiliation{Laboratoire de Physique des Solides, Univ. Paris-Sud, CNRS, UMR 8502, F-91405 Orsay Cedex, France}
\affiliation{Universit\'e de Sherbrooke, Sherbrooke, Qu\'ebec J1K 2R1, Canada}

\date{\today}

\begin{abstract}
The radiation generated by a quantum conductor should be correlated with electrons crossing it. We have measured the correlation between the fluctuations of the high frequency electromagnetic power and the low frequency transport in a tunnel junction. We have explored the regimes where electromagnetic fluctuations correspond to real photons and where they correspond to vacuum at very low temperature. We deduce from our data the intrinsic third moment of quantum shot noise, which appears to be frequency independent.
\end{abstract}

\pacs{72.70.+m, 42.50.Lc, 05.40.-a, 73.23.-b} \maketitle

\vspace{-0.5cm}

The existence of fluctuations of the electromagnetic field in vacuum even at zero temperature is a hallmark of the quantum nature of light. The quantized electromagnetic field at frequency $f$ is described by an harmonic oscillator. The uncertainty in the position of the oscillator in its ground state implies fluctuations of the electromagnetic field and an associated energy $\frac12hf$, the so-called half-photon of vacuum\cite{Loudon}. In a conductor, electromagnetic fluctuations (usually referred to as noise) are associated with fluctuations of local charges and currents, i.e. with the motion of electrons. Although these are described by a complex hamiltonian that includes e.g. the effects of disorder and electron-electron interactions, the zero-point motion of electrons leads to current fluctuations which, at equilibrium and zero temperature have a variance given by $S_{I^2}(f)=Ghf$ with $G$ the linear conductance of the sample \cite{Nyquist}. The factor $hf$ in this expression comes from the fact that high frequency current fluctuations in a conductor at very low temperature are indeed a direct consequence of the quantum nature of electricity.

Theoretical predictions in quantum optics are extremely powerful thanks to the fact that photon field operators in different modes commute with each other, and that the theory of photo-detection is well established \cite{Loudon,Glauber}. Quantum noise in mesoscopic conductors is a much younger field, and how to model the measurement process, which may involve macroscopic apparatus as well as other quantum conductors, is still a very active field of research. The main difficulty comes from the fact that current operators $\hat I$ (which belong to the Hilbert space of the electron gas) taken at different times or frequencies do not commute with each other. For example, the measurement of the classical current-current correlator $\langle I(f)I(-f)\rangle=\langle |I(f)|^2\rangle$ (with $I(f)$ the Fourier component at frequency $f$ of the current) i.e., the average power at frequency $f$, must be related with the quantum correlators $\langle \hat I(f)\hat I(-f)\rangle$ and $\langle \hat I(-f)\hat I(f)\rangle$. Those two differ by a quantity $Ghf$ and which one is measured is not a trivial question. These particular correlators of order two have however a clear physical picture in terms of photons: one corresponds to absorption of photons while the other one corresponds to emission \cite{Lesovik,Gavish,Aguado,Deblock}. Depending on the detector, one or the other can be measured, and the result $S_{I^2}(f)=Ghf$ is believed to describe the outcome of a classical detector\cite{Clerk}. In the case of higher order correlators, no such interpretation exists, and if there are some models of idealized detectors (such as a single spin \cite{Levitov}, a lossless $LC$ circuit read out by an electrometer \cite{Lesovik} or a 1D wire \cite{Bednorz}), these are far from real ones, diodes, mixers or digitizers.

Here we report an experiment that aims at deepen the understanding of quantum noise by the measurement of the correlator $C(f,\delta f)=\langle P_f(\delta f) V(-\delta f)\rangle$. $V(\delta f)$ denotes the voltage fluctuation at low frequency $\delta f$, $h\delta f\ll k_BT, eV$ with $V$ the dc voltage across the sample and $T$ the temperature. $P_f(\delta f)$ is the Fourier component at frequency $\delta f$ of the power fluctuations of the electromagnetic field oscillating at frequency $f$. $P_f(0)$ corresponds to the average power at frequency $f$, therefore it is proportional to the noise spectral density $S_{V^2}(f)=\langle V(f)V(-f)\rangle= \langle |V(f)|^2\rangle$. At finite frequency $\delta f$ this generalizes to $P_f(\delta f)\propto V(f)V(-f+\delta f)+V(-f)V(f+\delta f)$. Thus the correlator we measure is proportional to the third moment of voltage fluctuations $S_{V^3}(f,\delta f)=\langle V(\pm f)V(\mp f+\delta f)V(-\delta f)\rangle$, which describes the correlation between Fourier components of the voltage taken at three different frequencies. This quantity is closely related to the \emph{intrinsic} third moment of current fluctuations $S_{I^3}^{int}(f,\delta f)=\langle I(f)I(-f+\delta f)I(-\delta f)\rangle$ generated by the sample. Our experiment at the lowest temperatures corresponds to the quantum regime $hf\gg k_BT$. For $eV<hf$, no photon of frequency $f$ is emitted, and the noise detected at frequency $f$ is that of vacuum fluctuations. In this regime, we are measuring how these are correlated with the current itself.  We show in the following that $S_{I^3}^{int}(f,\delta f)=e^2I$ for a tunnel junction, with $I$ the dc bias current. This result, which corresponds to the quantum prediction provided we use the Keldysh prescription to order the current operators at different times, implies the high frequency zero point motion of electrons being correlated with the current itself. For the photon field in the coax cable connected to the sample, our result shows the existence of a discontinuity in the third order correlator of the electric field.

\begin{figure}
\begin{center}
\includegraphics[width=0.85\linewidth]{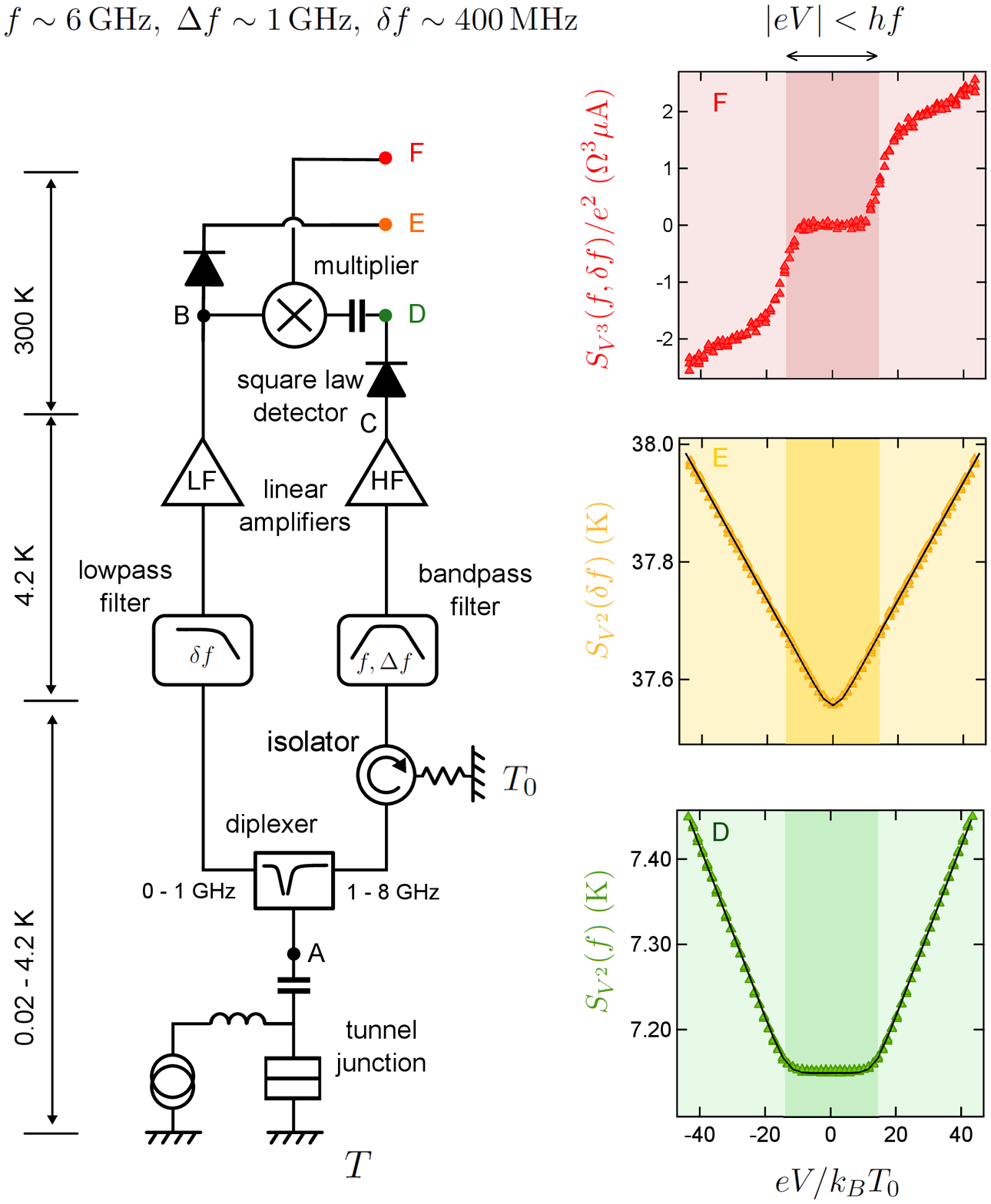}
\end{center}
\caption{(Color online). Experimental setup for the measurement of the third moment of voltage fluctuations $S_{V^3}(f,0)$. The symbol $\otimes$ represents a multiplier, which output is the product of its two inputs. The diode symbol represents a square law detector, which output is proportional to the low frequency part of the square of its input. \textit{Right}: All the measurements presented in these plots have been performed at $T=20\, \mathrm{mK}$. The solid lines in insets D and E corresponds to fits with Eq.(6) of appendix B. (D) (green triangles) Measured high frequency noise $S_{V^2}(f)$, with $f = 6 \, \mathrm{GHz}$, so that $hf/k_BT=15$. It exhibits a crossover between quantum noise  $|eV|<hf$ and shot noise $|eV|>hf$. (E) (orange triangles) Measured low frequency noise $S_{V^2}(\delta f)$, with $\delta f<400$ MHz, i.e. $h\delta f/k_BT<0.25$. It exhibits a crossover between thermal noise  $|eV|<k_BT$ and shot noise $|eV|>k_BT$. (F) (red triangle) Measured third moment of voltage fluctuations $S_{V^3}(f,\delta f)$.\label{fig1}}
\end{figure}

\vspace{0.2cm}\noindent\emph{Principle of the measurement.} We have chosen to perform the measurement on the simplest system that exhibits well understood shot noise, the tunnel junction. The sample is an Al/Al oxide/Al tunnel junction similar to that used for noise thermometry \cite{Lafe}. We apply a 0.1 T perpendicular magnetic field to turn the Al normal. The principle of the experimental setup, sketched in Fig. \ref{fig1}, is the following: the sample is biased by a dc current while ac voltage fluctuations across it (point A on Fig. \ref{fig1}) are amplified and detected in two frequency bands: a low frequency band (LF, left on Fig. \ref{fig1}) which provides at point B a voltage $v_B$ proportional to $V(\pm\delta f)$ for $|\delta f|<400$ MHz, and a high frequency band (RF, right on Fig. \ref{fig1}) centered at $f=6$ GHz, which provides at point C, $v_C\propto V(\pm f\pm\delta f')$ with $|\delta f'|<1$ GHz. A fast power detector (symbolized by a diode) measures $v_D(\delta f)\propto P_f(\delta f)$. Its dc output provides a measurement of the noise spectral density at frequency $f$, $\langle v_D\rangle\propto S_{V^2}(f)$, while its ac components are multiplied with $v_B$, the dc part of which gives $C(f,\delta f)\simeq C(f,0)$ (in the following we will replace $\delta f$ by 0 since $h\delta f\ll k_BT$). A more detailed description of the setup is given in appendix A.

\vspace{0.2cm}\noindent\emph{Results.} The measurement of the low frequency noise spectral density $S_{V^2}(0)$ shown on Fig. \ref{fig1}(E) allows us to extract the electron temperature $T$. Our lowest temperature is $T=20$mK. The curve of Fig. \ref{fig1}(D) shows the measurement of the high frequency noise spectral density $S_{V^2}(f)$, as obtained at point D. We observe a very clear plateau at low voltage which separates the quantum regime $|eV|<hf$ from the classical regime \cite{BuBlan}. The rounding of the curve at the threshold between the two regimes is due to the finite bandwidth $\Delta f_H$ as well as the finite temperature. In the quantum regime, the voltage fluctuations across the sample correspond to the vacuum fluctuations. To these adds the voltage noise of the HF amplifier, so that the noise temperature $R_0S_{I^2}/(2k_B)=7.1$ K (measured by $\langle v_D\rangle)$ at zero voltage is much higher than the equivalent noise temperature of the vacuum fluctuations, $hf/k_B=0.3$ K. Here $R_0=50\,\Omega$ is the characteristic impedance of the microwave circuitry, as well as the input impedance of the amplifiers. The resistance of the sample $R=50.4\,\Omega$ is close enough to $R_0$ so that corrections due to impedance mismatch can be neglected.

\begin{figure}
\begin{center}
\includegraphics[width=0.8\linewidth]{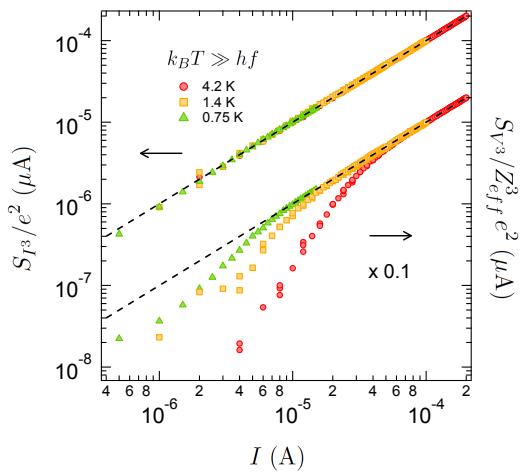}
\end{center}
\caption{(Color online). Normalized third moment of current and voltage fluctuations in the classical regime $hf\ll k_BT$ measured at different temperatures. \textit{Right axis}: third moment of voltage fluctuations $S_{V^3}(f,0)$ across the sample, that contain the environmental contributions. \textit{Left axis}: normalized intrinsic third moment of current fluctuations $S_{I^3}^{int}(f,0)$ obtained from $S_{V^3}(f,0)$ after subtraction of the environmental contributions. The data set have been multiplied by $0.1$ for clarity.\label{fig2}}
\end{figure}

Measurements of the third moment of current fluctuations are known to be strongly affected by environmental effects \cite{S3BR,Kindermann,Beenakker}. These could be avoided if we had ampmeters (i.e., amplifiers with very low input impedance) working at frequencies of several GHz and being noiseless. Unfortunately, such devices do not yet exist, and the best is to have a well controlled electromagnetic environment in order to be able to subtract environmental contributions with confidence. Environmental effects correspond to the back-action of the measuring apparatus, which cannot measure current without inducing voltage variations across the sample. These, in turn, influence electron transport. Voltage fluctuations are due to a) the noise generated by the amplifiers and b) the current fluctuations from the sample itself across the input impedance of the amplifiers $R_0$. We must precisely control and know these two parameters. In our setup everything, including the sample, is closely matched to $R_0$ and the noise generated by the HF amplifier is blocked by a circulator which load is well thermalized at the sample temperature. Thus the HF noise experienced by the sample is the thermal noise of that load. The noise emitted by the LF amplifier towards the sample is measured independently. Environmental contributions are well understood. They are due to the correlator $\langle P_{f_1}(f_2) V(-f_2)\rangle$. If the voltage across the sample does not fluctuate, $V(f_2)=0$, this correlator vanishes. The response of the noise of the sample measured at frequency $f_1$ to a small excitation at frequency $f_2$, i.e. $P_{f_1}(f_2)$, is given by $P_{f_1}(f_2)=\chi_{f_1}(f_2)V(f_2)$ where $\chi_{f_1}(f_2)$ is the noise susceptibility, which has been both calculated \cite{GR1} and measured \cite{GR2}. So what needs to be known are the voltage fluctuations $\langle |V(f_2)|^2\rangle=|Z_{eff}(f_2)|^2[S_{I^2}(f_2)+S_{I^2}^{env}(f_2)]$ where $Z_{eff}$ is the impedance equivalent to the sample in parallel with the amplifier, $S_{I^2}$ (resp. $S_{I^2}^{env}$) is the noise spectral density of the current fluctuations generated by the sample (resp. the environment, i.e. the amplifiers). The frequencies $(f_1,f_2)$ relevant for our experiment are $(f,\delta f)$, $(f,-f+\delta f)$ and $(\delta f, f)$, which physically correspond to both modulation of the HF noise by LF voltage fluctuations and modulation of LF noise by HF voltage fluctuations. Detailed expressions of the environmental corrections are given in appendix B. We show on Fig. \ref{fig2} the data in the classical regime (high temperature, $k_BT\gg hf$) for various temperatures. While environmental contributions depend on temperature, see the right part of Fig. \ref{fig2}, the intrinsic third moment is temperature independent, given by $S_{I^3}=e^2I$, see the left part of Fig. \ref{fig2}. This proves that environmental contributions are very well accounted for. It provides a strong confidence in their subtraction at the lowest temperatures, where the high frequency third moment is not known.

\begin{figure}
\begin{center}
\includegraphics[width=0.75\linewidth]{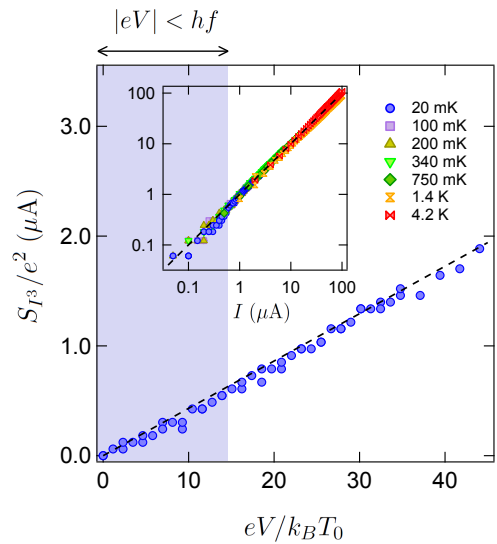}
\end{center}
\caption{(Color online). Measured normalized intrinsic third moment of current fluctuations $S_{I^3}^{int}(f,0)$ at the lowest temperature $T=20$ mK, as a function of the reduced voltage $eV/(k_BT)$. \textit{Inset}: $S_{I^3}^{int}(f,0)/e^2$ as a function of the dc current for seven temperatures between 20mK and 4.2K.\label{fig3}}
\end{figure}

We now focus on the data at the lowest temperature. Fig. \ref{fig1}F shows the third moment of the voltage fluctuations across the sample. We observe $S_{V^3}(f,0)=0$ at low voltage $eV<hf$. This depends on the noise temperature of the environment as well as its impedance. In ref. \cite{upon}, a similar measurement with a not so well controlled environment and less precision shows an example of a different result, because the environment is different. Here we obtain zero because the sample and the detection setup are impedance matched and because the influence of the noise of the amplifiers is weak enough. Once we subtract the environmental contributions, we obtain for the intrinsic third moment of current fluctuations $S_{I^3}^{int}(f,0)=e^2I$ over the whole voltage range, as shown in Fig. \ref{fig3}. There is no anomaly at $eV=hf$, in strong contrast with what happens for $S_{I^2}$, see Fig. \ref{fig1}D. This result is totally temperature independent over more than two decades of temperature, as demonstrated in the inset of Fig. \ref{fig3} where we report our data for all measured temperatures from 20mK to 4.2K.

\vspace{0.2cm}\noindent\emph{Discussion.} Previous measurements have shown that the third moment of current fluctuations created by a tunnel junction measured at low frequency,  is temperature independent and given by $S_{I^3}^{int}(0,0)=e^2I$ \cite{S3BR}. Here we have shown that  $S_{I^3}^{int}(f,0)=e^2I$ regardless of $f$. This coincides with what would happen if electrons crossing the tunnel barrier created randomly distributed, infinitely narrow  peaks of current, with a probability distribution given by the Poisson law. But in that case $S_{I^2}(f)$ would also be frequency independent. That the kink in $S_{I^2}(f)$ at $hf=eV$, the maximum frequency for photons being emitted by the junction, has no counterpart in $S_{I^3}^{int}(f,0)$ is very counterintuitive. But it means that the power of the current fluctuations at frequencies greater that $eV/h$, corresponding to the vacuum fluctuations, are correlated with the current fluctuations at low frequency. The difficulty of the interpretation of the third moment of current fluctuations comes from the fact that the quantum current operators $\hat I$ taken at different times or frequencies do not commute with each other. The various correlators involved in the calculation of $S_{I^3}^{int}(f,f')$ depend on voltage, frequency and temperature \cite{Salo}. Only certain combinations of those correlators give the result $S_{I^3}^{int}(f,f')=e^2I$. At least the so-called Keldysh ordering does \cite{Zaikin1, Zaikin2, FCSgeneral, FCScircuit}. But if we replace our HF detection (amplifier + square law detector) by a photomultiplier for example (which "clicks" only when it absorbs a real photon), what correlator would correctly predict the issue of the measurement is far from being clear and deserves theoretical investigations.

From measurements on a sample embedded in a microwave circuit we have calculated what would be its response in the absence of external circuit. This is usual in mesoscopic physics because the electromagnetic environment of a sample can be varied at will as opposed to the case of an atom in the vacuum, or even a cavity. It is however tempting to try to understand the physical meaning of our raw result for $S_{V^3}$ of Fig. \ref{fig1}(F): it represents the third moment of the fluctuations of the voltage (or the electric field) in the coax cable connected to the sample. In our setup, the sample is well matched to the coax cable, so the contribution from the amplifiers is small (see appendix B). Thus we have measured the intrinsic third moment of voltage fluctuations of a tunnel junction in the presence of a matched vacuum, i.e. connected to a semi-infinite coax cable of the same impedance. It appears to have a singular property: the amplitude of the fluctuating voltage, as given by $S_{V^2}$, is a continuous function of the bias voltage, both at zero and finite frequency. However, its third order correlator $S_{V^3}(f,0)$ is discontinuous at zero temperature and $V=hf/e$ (the width of this jump is set by the measurement bandwidth $\sim1$GHz and the temperature). It also means that as soon as photons of frequency $f$ are emitted, even with an infinitesimal flux, the electromagnetic field experiences an abrupt change, measured by the third moment. Again, a clear theoretical explanation is called for to understand this result.

\vspace{0.2cm}\noindent\emph{Conclusion.} Mesoscopic physics has started with the study of electron transport in quantum devices \cite{Imry,Datta}, followed by that of the noise (i.e., photons) generated by such devices \cite{BuBlan,Nazarov_book}. Our experiment provides the first observation of the correlation between them, i.e. between the electron transport at low frequency and the photon field at high frequency, both when photons are emitted ($eV>hf$) and when the electromagnetic field is solely due to vacuum fluctuations. In terms of the electrons only, our observations indicate that the intrinsic current fluctuations in a tunnel junction are given by $S_{I^3}^{int}(f,0)=e^2I$, regardless of the frequency. Thus the high frequency quantum current fluctuations are indeed correlated with their low frequency, classical counterpart. In terms of photons, we observe that the electromagnetic field exhibits a discontinuity at $eV=hf$ that shows up in a third order correlator and not in the second order ones.

\vspace{0.2cm}\noindent\emph{Acknowledgements.} We acknowledge fruitful discussions with M. Aprili, A. Bednorz, W. Belzig, M. B\"uttiker, A. Clerk, M. Devoret and B. Dou\c{c}ot. We thank B. Huard for lending us a cryogenic amplifier. This work was supported by ANR-11-JS04-006-01 and the Canada Excellence Research Chairs program.

\section{Appendix A: Experimental setup}

The tunnel junction is mounted on a rf sample holder placed on the mixing chamber of a dilution refrigerator, at temperature $T_0$ that we varied between 4.2K and 12mK. A bias tee, sketched in Fig. 1 by an inductor and a capacitor, allows to separate the dc current, imposed by a current source, from the current fluctuations, which can be read-out and amplified at point A. To select the frequencies at which we want to observe $S_{V^3}$, we use a diplexer that separates current fluctuations emerging from the sample into two branches depending on frequency: frequencies below 1 GHz (the LF branch, left on Fig. 1) and above 1 GHz (the HF branch, right on Fig. 1). Note that the diplexer is a non-dissipative element that does not add any noise. In the LF branch, the signal is low-pass filtered and amplified, to give a fluctuating voltage $v_B(t)$ which has Fourier components $v_B(\delta f)\propto V(\delta f)$ with $|\delta f|<\Delta f_L\simeq400$ MHz. In the RF branch, current fluctuations enter a circulator and a bandpass filter centered on the frequency $f=6$ GHz before being amplified by a 4-8 GHz cryogenic amplifier. Thus the voltage $v_C(t)$ has Fourier components $v_C(\pm f+\delta f')\propto V(\pm f+\delta f')$ with $|\delta f'|<\Delta f_H\simeq1$ GHz. A fast diode takes the square of $v_C(t)$ and filters out the high frequencies (close to $2f$), so that the voltage $v_D$ at its output has Fourier components given by $v_D(\delta f)\propto v_C(f'+\delta f)v_C(-f')$ with $f'=\pm f+\delta f'$ within the bandwidth of the HF branch, i.e. $|\delta f'|<\Delta f_{H}$. The dc output of the diode is given by :
\begin {equation}
\langle v_D \rangle\propto \int v_C(f+\delta f')v_C(-f-\delta f')d\delta f'\propto S_{V^2}(f)\Delta f_{H}
\label{eq:vD}
\end{equation}
where $S_{V^2}(f)=\langle V(f)V(-f)\rangle=\langle |V(f)|^2 \rangle$ is the \emph{time-average noise} (variance of the fluctuations) at frequency $f$. Experimental values of $v_D$ are reported in Fig. 1(D). In the right hand of Eq. (\ref{eq:vD}), we have neglected the frequency dependence of $S_{V^2}$ on the scale of $\Delta f_{H}$. However, solid line in Fig. 1(D) corresponds to a fit with Eq. (\ref{eq:formulas}) taking into account the frequency dependence of $S_{V^2}(f)$. Neglecting the frequency dependence just results in a slight overestimate of the temperature. Since the dc output of the diode $\langle v_D\rangle$ is the time-averaged noise power, its ac output $v_D(t)-\langle v_D\rangle$ has the physical meaning of \emph{fluctuations of the noise power}. This signal is (after further room temperature amplification, not shown), multiplied with the signal coming from the LF branch $v_B(t)$, using an analog, active multiplier. The dc output of the multiplier given by $\langle v_E\rangle\propto \langle v_B \times (v_D-\langle v_D\rangle)\rangle$ or, in terms of Fourier components:
\begin{equation}
\langle v_E\rangle \propto \int d\delta f d\delta f' v_B(-\delta f)v_C(f+\delta f'+\delta f)v_C(-f-\delta f')
\end{equation}
i.e. is proportional to $S_{V^3}(f,0)\Delta f_{L} \Delta f_{H}$.
 $\langle v_E\rangle$ is recorded as a function of the dc current in the sample and averaged for $\sim24$ hours for each temperature.

\section{Appendix B: Environmental contributions}

\begin{figure}
\begin{center}
\includegraphics[width=0.75\linewidth]{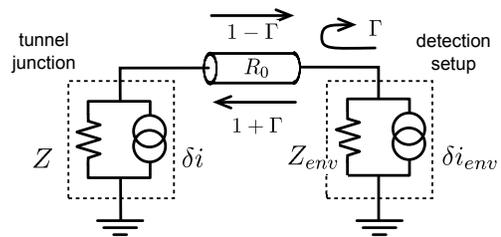}
\end{center}
\caption{Schematics of the detection setup equivalent to the high frequency part of the circuit depicted on Fig. 1. For clarity, the dc biasing circuit has been removed. $Z$ and $Z_{env}$ are the impedances of the sample and the environment, respectively. The current source $\delta i$ represents the fluctuating currents in the sample and $\delta i_{env}$ the current fluctuations emitted by the detection setup. $\Gamma=(Z-Z_{env})/(Z+Z_{env})$ is the voltage reflection coefficient of the sample.\label{figS1}}
\end{figure}

In this section, we describe the contributions that participate in the spectral density of the third moment of voltage fluctuations $S_{V^3}(f,0)=\langle V(f)V(-f)V(0) \rangle$. They are: the intrinsic third moment of current fluctuations $S_{I^3}^{int}(f,0)=\langle I(f)I(-f)I(0)\rangle$, the noise of the sample $S_{I^2}^{int}(f)$, the noise of the environment $S_{I^2}^{env}(f)$, the noise susceptibilities of the sample (response of the noise measured at low or high frequency to a low or high frequency excitation), the (complex) impedance of the sample $Z(f)$ and the impedance of the environment $Z_{env}\simeq 50\,\Omega$, see Fig. \ref{figS1}. Note that $V(f)$ and $I(f)$ refer to the Fourier component at frequency $f$ of voltage- and current fluctuations respectively, and $V(0)$ and $I(0)$ to voltage- and current fluctuations at low, but not strictly zero frequency.

\subsection{1. Theory}

\noindent The skewness of the current fluctuations in a conductor is strongly affected by its electromagnetic environment. An exhaustive experimental study of these environmental effects has been reported in [12] in the classical regime ($hf \ll k_BT$) and is in a perfect agreement with theoretical predictions [11,13]. The mechanism is that the noise at any frequency $f_2$ generated by the sample is modulated by the voltage fluctuations it experiences. The amplitude of such a modulation is given at low excitation frequency by $dS_2(f_2)/dV$. When the excitation frequency $f_1$ is not small, this quantity has to be replaced by the noise susceptibility $\chi_{f_1}(f_2) = \left\langle I(f_2)I(f_1-f_2) \right\rangle$ [14]. In the present experiment, we focus on the quantum regime $hf \gg k_BT, eV$ and the theory of environmental effects can be extended by introducing the noise susceptibility. We obtain:
\begin{equation}
S_{V^3}(f,0)=k [ - S_{I^3}^{int}(f,0)+ C_{env}^{fb}(f,0) + C_{env}^{ext}(f,0)]
\label{eq:contributions}
\end{equation}
\noindent where $k=Z_{eff}(0) \left|Z_{eff}(f)\right|^2$ and $Z_{eff}=ZZ_{env}/(Z+Z_{env})$. $C_{env}^{fb}(f,0)$ refers to the correlations induced by the feedback of the environment and $C_{env}^{ext}(f,0)$ to the correlations induced by the noise generated by the environment (thermal noise and/or noise of the amplifiers). These contributions to the total signal are shown in Fig. \ref{figS2}. They are given by (using $\chi_0(f)=\chi_f(0)$):

\begin{figure}
\begin{center}
\includegraphics[width=0.75\linewidth]{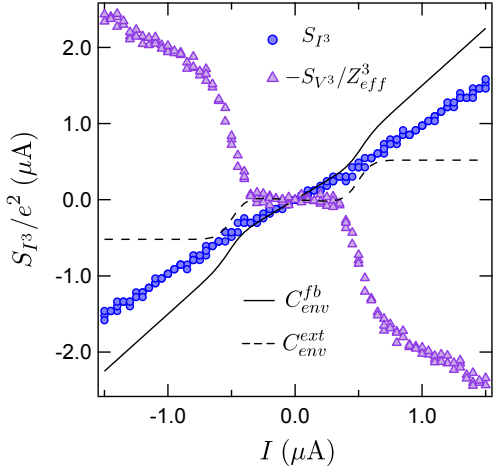}
\end{center}
\caption{Normalized $S_{V^3}(f,0)$ at $T=20 \, \mathrm{mK}$ (purple triangles) as a function of the dc current. $S_{I^3}^{int}(f,0)$ (blue circles) is obtained after subtraction of the environmental terms corresponding to the feedback of the environment $C_{env}^{fb}(f,0)$ (solid line) and the noise of the environment $C_{env}^{ext}(f,0)$ (dashed line), according to Eqs. (\ref{eq:contributions}) to (\ref{eq:formulas}).\label{figS2}}
\end{figure}

\begin{equation}
\begin{array}{ll}
C_{env}^{fb}(f,0)=&  Z_{eff}(0) \,S_{I^2}^{int}(0) \chi_{0}(f) \\
\\
& +2 \mathrm{Re} \left(Z_{eff}(f)\right) \, S_{I^2}^{int}(f) \chi_{f}(f)\\
\end{array}
\end{equation}
\begin{equation}
\begin{array}{ll}
C_{env}^{ext}(f,0)=& \frac{2Z_{eff}(0)\Gamma(0)}{1+\Gamma(0)} \,S_{I^2}^{env}(0) \chi_{0}(f)  \\
\\
&+ 2  \mathrm{Re}\left(\frac{2Z_{eff}(f)\Gamma(-f)}{1+\Gamma(-f)}\right) \, S_{I^2}^{env}(f) \chi_{f}(f)\\
\end{array}
\label{eq:Cenvext}
\end{equation}
\noindent where $\Gamma= \left(Z-Z_{env}\right)/\left(Z+Z_{env}\right)$ is the voltage reflection coefficient of the sample. The noise and noise susceptibilities of the tunnel junction at temperature $T$ depend on voltage and frequency according to:
\begin{equation}
\begin{array}{l}
S_{I^2}^{int}(f)=  \mathrm{Re}\left(\frac{k_BT}{2Z(f)}\right) \, \left\{F(eV+hf)+F(eV-hf)\right\}  \\
\\
\chi_{0}(f)= \mathrm{Re}\left(\frac{ek_BT}{2Z(f)}\right) \, \left\{F'(eV+hf)+F'(eV-hf) \right\} \\
\\
\chi_{f}(f)= \mathrm{Re}\left(\frac{ek_BT}{2Z(f)hf}\right) \, \left\{F(eV+hf)-F(eV-hf)\right\}
\end{array}\label{eq:formulas}
\end{equation}

\noindent with $F(\epsilon)=[1+\exp \epsilon/(k_BT)]^{-1}$ the Fermi-Dirac distribution. Note that $\chi_0(f)=\frac{dS_{I^2}(f)}{dV}$. In the classical regime $hf \ll k_BT$, $S_{I^2}(f)\simeq S_{I^2}(0)$,  $\chi_f(f)\simeq\chi_0(f)$ and we recover Eq. (1) of [13].

\begin{figure}
\begin{center}
\includegraphics[width=0.75\linewidth]{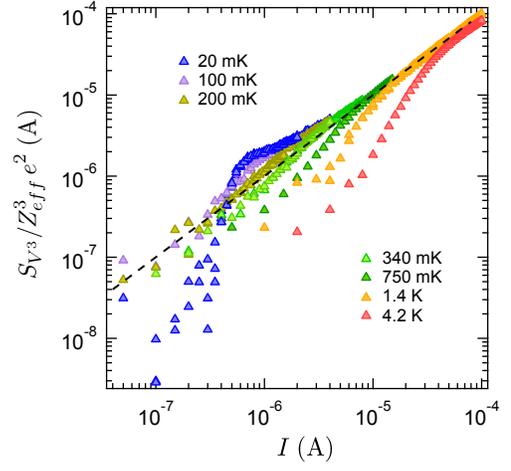}
\end{center}
\caption{Third moment of voltage fluctuations $S_{V^3}(f,0)$ across the sample as a function of the dc current, for different temperatures. Dashed line corresponds to the theoretical expectation in the classical regime: $S_{V^3}(0,0)=R_{eff}^3e^2 I$.\label{figS3}}
\end{figure}

\subsection{2. Experimental determination}

Here we describe the procedure for  the determination of the environmental terms in $S_{V^3}(f,0)$. We first consider the high voltage limit $eV\gg hf, k_BT$. We observe that $S_{V^3}(f,0)$ is linear in $I$ with a temperature-independent slope $\beta$, see Fig. \ref{figS3}. $\beta$ is independent of $C_{env}^{ext}$. It involves the overall gain of the setup as well as $Z$ and $Z_{env}$ at frequencies $f$ and 0. We have measured $Z$ vs. frequency with a network analyzer in the range $300 \, \mathrm{kHz} \, -  \, 8 \, \mathrm{GHz}$. We observe that the tunnel junction can be well modeled by a resistor $R \simeq 50.4 \, \Omega$ in parallel with a capacitor $C \simeq  \, 0.6\mathrm{pF}$ leading to a frequency cutoff of about 6 GHz. We take $Z_{env}=50\,\Omega$. Deviation from this value are irrelevant here and will be discussed later. Thus the high voltage slope of $S_{V^3}$ gives us the gain of the setup. In order to focus on the low voltage part, we define:
\begin{equation}
\alpha(I)=(S_{V^3}-\beta I)/(k e^2)
\label{eq:alpha}
\end{equation}

\begin{figure}
\begin{center}
\includegraphics[width=0.75\linewidth]{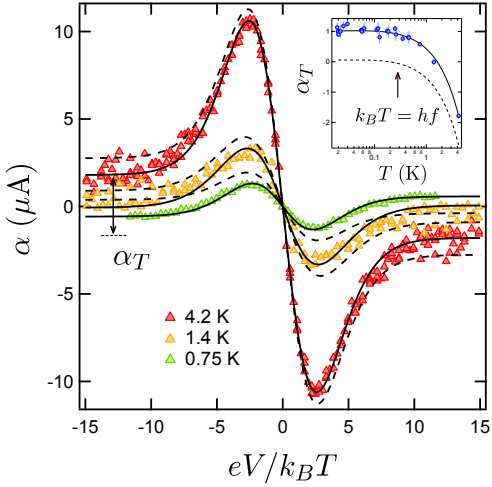}
\end{center}
\caption{$\alpha$ as a function of $eV/(k_BT)$ at different temperatures in the classical regime $k_BT \gg hf$, from 0.75K to 4.2K, calculated according to Eq. (\ref{eq:alpha}) from the measurement of $S_{V^3}(f,0)$. Dashed lines are expectations for a $Z_{env}=R_0=50\, \Omega$ environmental impedance. Solid lines correspond to $Z_{env}=48\,\Omega$. Inset: temperature dependence of the high current value $\alpha_0$ of $\alpha$. Solid line is the expectation for $Z_{env}=48\,\Omega$, given by Eq. (\ref{eq:alpha0}).\label{figS4}}
\end{figure}

\noindent which is plotted on Fig. \ref{figS4}. To extract the external noise term we need to know the noise of the environment. The low frequency noise $S_{I^2}^{env}(0)$ is measured \textit{in situ} by turning the junction superconducting (i.e., at zero magnetic field). Then the low frequency noise of the environment is fully reflected by the sample and measured at point E, on Fig. 1. The gain of the LF branch alone is calibrated by the measurement of $S_{I^2}(0)$. We obtain $S_{I^2}^{env}(0)= 2k_BT_N^{LF}/R_0$ with $T_N^{LF}=11.6 \pm 0.1\, \mathrm{K}$. The high frequency noise is that of the load of the circulator at temperature $T$,  plus the noise of the HF amplifier that leaks through the circulator, which corresponds to a thermal noise of $T_H=10$mK (obtained with a measurement similar to that of the LF branch):
\begin{equation}
S_{I^2}^{env}(f)=[hf \coth \frac{hf}{2k_BT}+2k_BT_H]/R_0
\label{eq:S2env_f}
\end{equation}
\noindent The contribution of the external noise is proportional to the reflection coefficient times the noise of the environment, see Eq. (\ref{eq:contributions}). At low frequency $\Gamma(0)$ is very small but $S_{I^2}^{env}(0)$ is large whereas at $f=6$ GHz, $\Gamma(f)$ is larger but $S_{I^2}^{env}(f)$ tiny. We show on Fig. \ref{figS4} in dashed lines the expectations for $\alpha(I)$ taking the measured values of the noise and impedances. The fit is good at 4.2K but the overall amplitude of $\alpha(I)$ is not perfectly well reproduced at lower temperature. This is due to the deviation of the input impedance of the amplifiers from their nominal value $R_0=50\;\Omega$. In order to evaluate this effect we plot on the inset of \ref{figS4} the high current asymptotic value $\alpha_0$ of $\alpha$ as a function of temperature,
\begin{equation}
\alpha_0=\alpha(eV\gg k_BT)=\gamma(0)S_{I^2}^{env}(0)+\gamma(f)S_{I^2}^{env}(f)
\label{eq:alpha0}
\end{equation}
\noindent where $\gamma(0)$ (resp. $\gamma(f)$) is a temperature-independent coefficient which depend on the impedances at frequency 0 (resp. $f$), see Eq. (\ref{eq:Cenvext}). Thus the low frequency environmental noise contributes only to a constant in $\alpha_0$. Since it is large, a very precise knowledge of $\gamma(0)$ is required, beyond the precision we could reach (moreover the input impedance of the LF amplifier has a strong frequency dependence over the almost two decades frequency window it covers). Also we have prefered to determine $\gamma(0)$ from the measurement of $\alpha_0(T)$. Dashed curves in Fig. \ref{figS4} and its inset correspond to calculated values of $\gamma(0)$ and $\gamma(T)$ taking $Z_{env}=R_0$. Slightly increasing the value of $\gamma(0)$ to fit the 4.2K data results in a much better fit of both the temperature dependence of $\alpha_0$ and the current dependence of $\alpha$, see solid lines in Fig. \ref{figS4} and its inset. It corresponds to $Z_{env}=48\,\Omega$, a very plausible value.

\end{document}